\documentclass{aa}
\usepackage{graphics}

\def\ref{\par \noindent \hang}

\def\approxlt{\mathrel{\hbox{ \lower .5ex \hbox {$\sim$}
	\llap{\raise .15 ex \hbox{$<$}} }}}
\def\approxgt{\mathrel{\hbox{ \lower .5ex \hbox {$\sim$}
	\llap{\raise .15 ex \hbox{$>$}} }}}

\def\multleft#1{\hbox to size{\vbox {\halign {\lft{##}\cr #1}}\hfill}\par}
\def\multright#1{\hbox to size{\vbox {\halign {\rt{##}\cr #1}}\hfill}\par}

\def\today{\ifcase\month\or January\or February\or March\or April\or May\or
      June\or July\or August\or September\or October\or November\or December\fi
      \space\number\day, \number\year}
\def\<{\thinspace}
\def\boxit#1{\vbox{\hrule\hbox{\vrule\kern3pt\vbox{\kern3pt
	  #1 \kern3pt}\kern3pt\vrule}\hrule}}
\def\H2{\hbox{H$_{2}$~}}

\def\ref{\par\noindent\hangindent 20 pt}

\def\ltsim{\raise 2pt \hbox {$<$} \kern-1.1em \lower 4pt \hbox {$\sim$}}
\def\gtsim{\raise 2pt \hbox {$>$} \kern-1.1em \lower 4pt \hbox {$\sim$}}

\begin{document}
 
\input psfig.sty
\thesaurus{03(11.01.2; 11.06.2; 11.16.1)}

\title{Optical surface photometry of radio galaxies - II. Observations and data analysis}

\author{Federica Govoni \inst{1, 3}, Renato Falomo\inst{1}, Giovanni Fasano\inst{1} and Riccardo Scarpa\inst{2}}
\offprints{F. Govoni (fgovoni@astbo1.bo.cnr.it)\\
Based on observations collected at the European Southern Observatory,
La Silla, Chile, and at the Nordic Optical Telescope, La Palma.}

\institute{Osservatorio Astronomico di Padova, vicolo dell'Osservatorio 5, I--35122 
Padova, Italy
\and
Space Telescope Science Institute, 3700 San Martin Drive, Baltimore, MD 
21218, U.S.A
\and
Dip. di Astronomia, Univ. di Bologna, Via Ranzani 1,I--40127 Bologna, Italy}
\date{Received; Accepted}

\maketitle

\markboth{F. Govoni et al.: Optical surface photometry of radio galaxies - II}
{F. Govoni et al.: Optical surface photometry of radio galaxies - II}

\begin{abstract}
Optical imaging observations for 50 radio galaxies are presented. 
For each object isophotal contours, photometric profiles, structural
parameters (position angle, ellipticity, Fourier coefficients), and
total magnitudes are given. These observations, obtained in the Cousins R
band, complement the data presented in a previous paper and are part
of a larger project aimed at studying the optical properties of low
redshift ($z\leq 0.12$) radio galaxies (Govoni et al. 1999). 
Comments for each individual source are reported.


\end{abstract}

\section{Introduction}
Only a small fraction of elliptical galaxies emit at radio
wavelengths.  This is probably due to the combination of two factors:
1) the
lifetime of the radio source is much less than the typical lifetime of
the host galaxy; 2) not all elliptical galaxies may harbor the
conditions for radio activity. Therefore, a detailed analysis and comparison of
the properties of radio and non-radio galaxies are of great relevance 
for understanding the phenomenon.
Previous optical studies (Hine \& Longair 1979; Longair \&
Seldner 1979; Lilly \& Prestage 1987; Prestage \& Peacock 1988; Owen
\& Laing 1989; Smith \& Heckman 1989a, b;
 Owen \& White 1991; Gonzalez-Serrano et al. 1993;
de Juan et al. 1994; Colina \& de Juan 1995; Ledlow \& Owen 1995) have
investigated the role of galaxy-galaxy interaction in
the creation and fueling of nuclear radio sources, as well as the connection 
between overall galaxy properties (e.g. luminosity and scale length) and
radio morphology.

In order to contribute to the study of the photometrical and morphological
properties of the galaxies hosting radio sources,
we have undertaken a systematic study of low redshift radio galaxies
in the optical band.

In a previous paper (Fasano et al. 1996; hereafter Paper I)
we presented structural (position angle, ellipticity, Fourier
coefficients) and photometric profiles together with isophotal
contours in the B and R bands for 29 galaxies extracted from
a complete sample of 
95 radio galaxies in the redshift range $0.01\leq z \leq0.12$.

 Here, we give
the results for 50 more galaxies observed in the Cousins R band,
bringing to 79, 83\% of the original sample, the total number of sources 
for which we were able to secure data of the required quality.
A full discussion of the astrophysical implications of these
observations is given in Govoni et al. (1999), where
the overall properties of galaxies hosting
radio emission are compared with radio quiet ellipticals.

%-------------------TABLE 1-------------------------------------

\begin{table*}
\begin{center}
\begin{tabular}{cccccccccc} 

\multicolumn{10}{c}{{\bf Table 1.} The sample}\\
\multicolumn{10}{c}{}\\
\hline\\

\multicolumn{1}{c}{IAU name} & \multicolumn{1}{c}{Other name} & \multicolumn{1}{c}{Smp.} & \multicolumn{1}{c}{RA(2000)} & \multicolumn{1}{c}{DEC(2000)} & \multicolumn{1}{c}{z} & \multicolumn{1}{c} {K$_R$} & \multicolumn{1}{c}{A$_V$}& \multicolumn{1}{c}{FR
} & \multicolumn{1}{c}{Ref.$^a$} \\
\medskip  (1)  & (2) & (3) & (4) & (5) & (6) & (7) & (8) & (9) & (10)\\
\hline\\
$0255+058$  & 3C 75        & WP &   02 57 41.5  &   $+$06 01 37     & 0.023  &0.02  &0.54&    I &   1   \\
$0257-398$  &              & EK &   02 59 26.6  &   $-$39 40 37     & 0.066  &0.06  &0.10&    II&   2   \\
$0307-305$  &              & EK &   03 10 00.9  &   $-$30 19 33     & 0.066  &0.06  &0.10&    II&   2   \\
$0312-343$  &              & EK &   03 14 32.7  &   $-$34 07 40     & 0.067  &0.06  &0.13&    I &   2   \\
$0325+023$  & 3C 88        & WP &   03 27 53.9  &   $+$02 33 42     & 0.030  &0.03  &0.51&   II &   1   \\
$0332-391$  &              & EK &   03 34 07.2  &   $-$39 00 04     & 0.063  &0.05  &0.10&    I &  2,3  \\
$0344-345$  &              & EK &   03 46 30.5  &   $-$34 22 47     & 0.053  &0.04  &0.06&    I &  2,3  \\
$0427-539$  &              & WP &   04 29 07.3  &   $-$53 49 40     & 0.038  &0.03  &0.17&    I &   1   \\
$0430+052$  & 3C 120       & WP &   04 33 10.5  &   $+$05 22 15     & 0.033  &0.03  &0.67&    I &   1   \\
$0434-225$  & OF-257       & EK &   04 36 35.9  &   $-$22 26 34     & 0.069  &0.06  &0.19&    I &   2   \\
$0446-206$  & OF-277.2     & EK &   04 48 29.9  &   $-$20 32 13     & 0.073  &0.06  &0.19&    I &   2   \\
$0452-190$  &              & EK &   04 54 17.7  &   $-$18 55 29     & 0.039  &0.03  &0.26&    I &   2   \\
$0453-206$  & NGC 1692     & WP &   04 55 23.6  &   $-$20 34 15     & 0.035  &0.03  &0.19& I/II &   1   \\
$0511-305$  &              & EK &   05 13 31.2  &   $-$30 28 49     & 0.058  &0.05  &0.10&   II &  2,3  \\
$0533-377$  &              & EK &   05 35 22.3  &   $-$37 43 13     & 0.096  &0.09  &0.16&    I &   2   \\
$0546-329$  &              & EK &   05 48 27.4  &   $-$32 58 37     & 0.037  &0.03  &0.16&    I &  2,3  \\
$0548-317$  &              & EK &   05 50 49.4  &   $-$31 44 26     & 0.034  &0.03  &0.13&   II &   2   \\
$0620-526$  &              & WP &   06 21 42.4  &   $-$52 41 32     & 0.051  &0.04  &0.30&    I &   1   \\
$0625-354$  & OH-342       & WP &   06 27 06.4  &   $-$35 29 16     & 0.055  &0.04  &0.38&    I &   2   \\
$0625-536$  &              & WP &   06 26 20.4  &   $-$53 41 35     & 0.054  &0.05  &0.34&   II &   1   \\
$0634-205$  &              & EK &   06 36 32.9  &   $-$20 34 53     & 0.056  &0.05  &1.22& I/II &   2   \\
$0712-349$  &              & EK &   07 13 56.3  &   $-$35 01 26     & 0.044  &0.03  &1.12&    I &   2   \\
$0718-340$  &              & EK &   07 20 47.2  &   $-$34 07 05     & 0.029  &0.02  &1.28&I/ II &   2   \\
$0806-103$  & 3C 195       & WP &   08 08 53.3  &   $-$10 27 40     & 0.110  &0.10  &0.58&   II &   1   \\
$0915-118$  & Hyd A, 3C 218& WP &   09 18 05.5  &   $-$12 05 43     & 0.054  &0.04  &0.29&    I &   1   \\
$0940-304$  &              & EK &   09 42 23.2  &   $-$30 44 11     & 0.038  &0.03  &0.58&    I &   2   \\
$0945+076$  & 3C 227       & WP &   09 47 45.6  &   $+$07 25 20     & 0.086  &0.08  &0.19&   II &   1   \\
$1002-320$  &              & EK &   10 04 39.3  &   $-$32 16 43     & 0.089  &0.08  &0.51&    I &  2,3  \\
$1043-290$  &              & EK &   10 46 09.7  &   $-$29 21 10     & 0.060  &0.05  &0.32&    I &   2   \\
$1053-282$  & OL-288       & EK &   10 55 32.5  &   $-$28 31 33     & 0.061  &0.05  &0.32&    I &   2   \\
$1056-360$  &              & EK &   10 58 54.6  &   $-$36 19 20     & 0.070  &0.06  &0.32& I/II &  2,3  \\
$1107-372$  & NGC 3557     & EK &   11 09 57.4  &   $-$37 32 17     & 0.010  &0.01  &0.51&    I & 2     \\
$1123-351$  &              & EK &   11 25 52.5  &   $-$35 23 40     & 0.032  &0.03  &0.42&    I & 2     \\
$1251-122$  & 3C 278       & WP &   12 54 35.1  &   $-$12 34 07     & 0.015  &0.02  &0.22&    I & 1     \\
$1251-289$  &              & EK &   12 54 40.3  &   $-$29 13 39     & 0.057  &0.05  &0.38&    I & 2     \\
$1257-253$  &              & EK &   12 59 48.8  &   $-$25 39 39     & 0.065  &0.06  &0.42&    I & 2     \\
$1258-321$  &              & EK &   13 01 00.6  &   $-$32 26 28     & 0.017  &0.02  &0.38&    I & 2     \\
$1318-434$  & NGC 5090      & WP &   13 21 12.4  &   $-$43 42 15     & 0.011  &0.01  &0.53&    I & 1     \\
$1323-271$  &              & EK &   13 26 10.2  &   $-$27 25 36     & 0.044  &0.03  &0.32&    U & 2     \\
$1333-337$  & IC4296       & WP &   13 36 38.8  &   $-$33 57 56     & 0.013  &0.01  &0.26& I/II & 1,2,3     \\
$1344-241$  &              & EK &   13 47 12.1  &   $-$24 22 21     & 0.020  &0.02  &0.35&    I & 2     \\
$1354-251$  &              & EK &   13 57 16.0  &   $-$25 23 23     & 0.038  &0.03  &0.35&    I & 2     \\
$1400-337$  & NGC 5419      & EK &   14 03 38.5  &   $-$33 58 43     & 0.014  &0.01  &0.35&    I & 2     \\
$1404-267$  & IC4374       & EK &   14 07 29.4  &   $-$27 01 02     & 0.022  &0.02  &0.32&    I?& 2     \\
$1637-771$  & PGC 058874   & WP &   16 44 15.7  &   $-$77 15 48     & 0.041  &0.03  &0.47&   II & 1,3     \\
$1717-009$  & 3C 353       & WP &   17 20 27.8  &   $-$00 58 46     & 0.031  &0.03  &0.61&   II & 1     \\
$1733-565$  &              & WP &   17 37 35.3  &   $-$56 34 03     & 0.098  &0.09  &0.47&   II & 1     \\
$1928-340$  & PGC 063333   & EK &   19 31 37.3  &   $-$33 54 40     & 0.098  &0.09  &0.45&    I & 2     \\
$1949+023$  & 3C 403       & WP &   19 52 16.2  &   $+$02 30 23     & 0.059  &0.05  &0.83&   II & 1     \\
$2221-023$  & 3C 445       & WP &   22 23 49.8  &   $-$02 06 13     & 0.057  &0.05  &0.32&   II & 1     \\
\hline\\

\multicolumn{7}{l}{\scriptsize $^a$ Radio classification derived from :}\\
\multicolumn{7}{l}{\scriptsize (1) Morganti et al. (1993) or Morganti et al. (1997);}\\
\multicolumn{7}{l}{\scriptsize (2) Ekers et al. (1989);}\\
\multicolumn{7}{l}{\scriptsize (3) Jones \& McAdam (1992).}\\

\end{tabular}
\end{center}
\end{table*}

%----------TABLE 2--------------------------------------------------

\begin{table*}
\begin{center}
\begin{tabular}{cllc} 
\multicolumn{4}{c}{{\bf Table 2.} Observation run of the sample of radio galaxies}\\
\multicolumn{4}{c}{}\\
\hline\\
\multicolumn{1}{c}{Run} & \multicolumn{1}{c}{Date} & 
\multicolumn{1}{c}{Instrumentation} & \multicolumn{1}{c}{CCD scale}\\
  & 	&  & ($\arcsec$/pixel) \\
\hline\\
1  &1993 July 19-21         & ESO 2.2m + EFOSC2     & 0.332 \\
2  &1994 September 10-11    & ESO 2.2m + EFOSC2     & 0.332 \\
3  &1996 February 9-11      & ESO 1.5m Danish+DFOSC & 0.403 \\
4  &1998 March 30-April 2   & ESO 1.5m Danish+DFOSC & 0.403 \\
5  &1998 September 21-24    & NOT-HIRAC             & 0.110 \\
\hline\\
\end{tabular}\\
\end{center}
\end{table*}

%-----------------------------------------------------------------

\section{Observations and data analysis}

\subsection{The sample}

The sample is composed of radio galaxies in the redshift range
$0.01\leq z \leq 0.12$ extracted from two complete surveys of radio sources.

The first one is the all sky survey of radio sources with radio flux
at 2.7 GHz greater than 2 Jy by Wall \& Peacock (1985; hereafter WP). From
this survey we extracted all objects, in the above redshift range,
classified as radio galaxies (see
also Tadhunter et al. 1993) at declination $\delta<10^\circ$. 

The second list is the Ekers et al. (1989; hereafter EK) catalogue of
radio galaxies with flux at 2.7 GHz greater than 0.25 Jy and $m_b<17.0$, 
in the declination zone $-40^\circ<\delta<-17^\circ$.
All objects classified as E or S0 in the above redshift range were included
in our list. Basic data for the 50 objects presented in this paper are
summarized in Table 1. Columns 1 and 2 give for each object
IAU and other names, column 3 gives the subsample, columns 4 and
5 the equatorial coordinates (equinox 2000.0), and column 6 the redshift.
Columns 7 and 8 report the K-correction in the R band and the galactic
extinction in the V band.
We derived the galactic extinction interpolating the data for galactic 
hydrogen column density given by Stark et al. (1992) and assuming 
$A_V/E_{B-V}=R=3.2$ and $E_{B-V}=N_H/5.1 \times 10^{21}$ (Knapp \& Kerr 1974).
In column 9 and 10 we give the radio classification and reference.

Based on the radio morphology, sources were divided into FRI and FRII 
radio classes following the Fanaroff and Riley scheme 
(Fanaroff \& Riley 1974).
Most sources in the WP sample were imaged by Morganti et al. (1993)
with the Very Large Array (VLA) and the Australia Telescope Compact
Array (ATCA), while VLA radio images are available for most
objects in the EK sample. Objects with transitional properties
or with unclear classification are marked as I/II, while
a U marks the only unresolved source $1323-271$.
For IC4374 (labeled with a question mark) we were not able to find a 
radio image, thus we include it in the FRI class because its radio luminosity
at 178 MHz is below $2\times10^{25}$ Watt/Hz, the dividing luminosity
between the two classes.

\subsection {Observations}

Observations were secured in five different observing runs.  Beside
three galaxies ($0307-305$, $0332-391$ and $1928-340$) included here, the
results of the imaging in B and R bands obtained with the ESO-2.2\ m
telescope during runs 1 and 2 (Table 2) have been reported in Paper I.
Data presented here were obtained in three more observing runs (run 3,
4, 5) with either the ESO-Danish 1.5\ m telescope or the Nordic
Optical Telescope (NOT). The journal of observations is given in
Table 3, where for each object we report the run of observation, the
total integration time, the atmospheric seeing expressed by the full
width half maximum (FWHM) of stellar images, and the sky surface
brightness, together with its estimated 1$\sigma$ uncertainty. 
The galaxy total apparent R band magnitude (corrected
for galactic extinction) computed by extrapolating to 
infinity the surface brightness profile is also given.
This value does not include the K-correction.

For most objects a short ($\sim$ 2 min.) and a long exposure (Table 3)
were obtained, so we also have an unsaturated image of the nuclear
region. In a few cases, the presence of bright stars in the field
forced us to take several short exposures, subsequently combined to
form a final, deep image.
Photometric conditions were generally good during the observations,
as confirmed by repeated observations of photometric standard stars
selected from the Landolt (1992) list. Comparison of  
the photometric zero point  for different nights indicates 
an average internal photometric accuracy of 5$-$10\%. This, combined with
the small uncertainty on the sky surface brightness (1$-$2\%),
gives a global internal photometric accuracy of the order of 10\%.
The atmospheric seeing was generally around $1$ arcsec, and the CCD
pixel size (Table 2) were always sufficiently small to ensure proper sampling
of the telescope point spread function (PSF).

% ---- Table 3 ----------------------------------------------------

\begin{table*}
\begin{center}
\begin{tabular}{ccrcccr}
\multicolumn{7}{c}{{\bf Table 3.} Journal of observations}\\
\multicolumn{7}{c}{}\\
\hline\\
\multicolumn{1}{c}{IAU name} &\multicolumn{1}{c}{run} &\multicolumn{1}{c} {exp} &\multicolumn{1}{c}{FWHM} &\multicolumn{1}{c}{ $\mu_{sky}$}&\multicolumn{1}{c}{ $\sigma_{sky}$}&\multicolumn{1}{c}{ m$_{tot}$} \\
\medskip   &   & (sec) & (arcsec) & (mag/arcsec$^2$)& (mag/arcsec$^2$)& (mag) \\
\hline\\
 $0255+058$ &5 &900  & 1.02 &  20.7 &   0.019 &12.75 \\    
 $0257-398$ &3 &1200 & 1.33 &  21.2 &   0.005 &14.38 \\   
 $0307-305$ &2 &900  & 1.40 &  20.5 &   0.007 &14.70 \\   
 $0312-343$ &3 &1200 & 1.29 &  21.3 &   0.006 &14.01 \\   
 $0325+023$ &5 &900  & 1.21 &  20.8 &   0.012 &12.70 \\   
 $0332-391$ &2 &1200 & 1.70 &  20.7 &   0.007 &14.06 \\   
 $0344-345$ &3 &1500 & 1.44 &  21.1 &   0.004 &14.70 \\   
 $0427-539$ &3 &600  & 1.13 &  21.2 &   0.005 &12.69 \\ 
 $0430+052$ &3 &1200 & 1.33 &  20.8 &   0.004 &13.01 \\ 
 $0434-225$ &3 &1200 & 1.33 &  21.3 &   0.007 &13.33 \\   
 $0446-206$ &3 &1200 & 1.16 &  21.1 &   0.005 &14.81 \\   
 $0452-190$ &3 &1200 & 1.35 &  21.4 &   0.009 &13.00 \\   
 $0453-206$ &3 &600  & 1.37 &  21.4 &   0.007 &12.57 \\   
 $0511-305$ &3 &1800 & 1.39 &  21.2 &   0.004 &14.66 \\   
 $0533-377$ &3 &1800 & 1.49 &  21.4 &   0.009 &14.57 \\   
 $0546-329$ &3 &1200 & 1.27 &  21.3 &   0.006 &12.34 \\   
 $0548-317$ &3 &900  & 1.50 &  20.8 &   0.010 &13.42 \\   
 $0620-526$ &3 &600  & 1.64 &  20.8 &   0.017 &12.53 \\   
 $0625-354$ &3 &900  & 1.25 &  20.0 &   0.009 &13.11 \\   
 $0625-536$ &3 &1200 & 1.21 &  20.9 &   0.006 &12.39 \\   
 $0634-205$ &3 &1800 & 1.31 &  20.9 &   0.008 &13.90 \\   
 $0712-349$ &3 &1200 & 1.30 &  20.2 &   0.010 &12.98 \\   
 $0718-340$ &3 &1200 & 1.25 &  20.9 &   0.008 &12.15 \\   
 $0806-103$ &3 &600  & 1.20 &  20.8 &   0.007 &15.20 \\   
 $0915-118$ &3 &600  & 1.35 &  20.1 &   0.009 &13.23 \\   
 $0940-304$ &3 &600  & 1.20 &  20.7 &   0.013 &13.12 \\   
 $0945+076$ &3 &1800 & 1.23 &  20.4 &   0.010 &15.19 \\
 $1002-320$ &3 &1200 & 1.17 &  19.9 &   0.008 &14.52 \\
 $1043-290$ &3 &300  & 1.17 &  20.5 &   0.012 &13.22 \\
 $1053-282$ &3 &1800 & 1.13 &  19.7 &   0.006 &13.86 \\
 $1056-360$ &3 &1800 & 1.19 &  19.6 &   0.006 &14.64 \\
 $1107-372$ &3 &300  & 1.2  &  19.5 &   0.005 &9.53  \\
 $1123-351$ &3 &510  & 1.21 &  20.3 &   0.010 &11.92 \\
 $1251-122$ &3 &600  & 1.25 &  20.1 &   0.009 &10.63 \\
 $1251-289$ &3 &1800 & 1.17 &  18.9 &   0.004 &12.41 \\
 $1257-253$ &3 &1200 & 1.2  &  19.8 &   0.006 &13.92 \\
 $1258-321$ &3 &1200 & 1.10 &  19.8 &   0.015 &10.71 \\
 $1318-434$ &3 &600  & 1.20 &  19.9 &   0.012 &9.98  \\
 $1323-271$ &3 &1800 & 1.10 &  19.5 &   0.013 &13.14 \\
 $1333-337$ &3 &600  & 1.20 &  19.9 &   0.006 &9.97  \\
 $1344-241$ &3 &1200 & 1.25 &  20.4 &   0.010 &12.37 \\
 $1354-251$ &3 &600  & 1.25 &  20.4 &   0.007 &13.36 \\
 $1400-337$ &4 &1200 & 1.25 &  20.9 &   0.010 &9.72  \\
 $1404-267$ &4 &1200 & 1.25 &  20.8 &   0.012 &11.71 \\
 $1637-771$ &4 &1200 & 1.55 &  20.6 &   0.008 &13.54 \\
 $1717-009$ &4 &1200 & 1.15 &  20.7 &   0.010 &13.87 \\
 $1733-565$ &4 &900  & 1.25 &  20.2 &   0.008 &15.22 \\
 $1928-340$ &2 &1200 & 1.20 &  20.2 &   0.018 &14.31 \\
 $1949+023$ &5 &1200 & 0.72 &  20.7 &   0.010 &14.11 \\
 $2221-023$ &5 &1200 & 0.67 &  20.9 &   0.012 &14.46 \\
\hline\\
\end{tabular}\\
\end{center}
\end{table*}

%\normalsize

\subsection{Data reduction and surface photometry}

Data reduction is extensively described in Paper I.
Here we simply remind the reader
that the IRAF-ccdred package was used for the basic reduction
(bias subtraction, image trimming, flat fielding, cosmic rays, etc.).
The dark current turned out to be insignificant and was neglected.
After flat fielding, images were characterized by a quite regular 
sky background, well fitted by a first order polynomial. 

Final images are shown in Fig. 1, where it is seen that 
selected sources cover an area of hundreds of arc-seconds square,
ideal for two-dimensional isophotal analysis, 
and can be traced down to a surface brightness of
$\mu_R\sim 25$~mag/arcsec$^2$.
It is also evident that these radio galaxies are often observed 
in highly crowded regions, with stars and/or nearby galaxies projected
on-top. Sky subtraction and isophotal analysis
(drawing, cleaning and fitting) was performed using the AIAP package
(Fasano 1990), which due to its high degree of interactivity, 
is particularly suitable in analyzing the morphology of
galaxies embedded in such high density regions.

The problem of obtaining reliable surface photometry of dumbbell
systems was faced by adopting the {\it two-galaxy} fitting strategy
outlined in Paper I, which allows us to fully separate the two galaxies. 
Contour plots for all dumbbell systems, together with those of the two
members are shown in Fig. 2.

From this analysis we derived photometric and structural parameters (surface
brightness, ellipticity, position angle and Fourier coefficients)
as a function of the equivalent
radius  $r = a\times(1-\epsilon)^{1/2}$ where $a$ is the semimajor axis and 
$\epsilon$ is the ellipticity of the ellipse fitting a given isophote.
Isophotes can not be fitted in the innermost few arcsec of the galaxy
because of the small number of pixels involved.
To cope with this limitation, we extracted an azimuthally averaged radial
profile, centered on the center of the first useful isophote.
If the nucleus was saturated, the short exposure was used.
The agreement between this average radial profile and that obtained
from isophote fitting  was always
excellent in the common region, thus the two profiles were joined smoothly
to fully model the $core$ and the outer region of the galaxies. 
In the following analysis we consider this combined profile as 
the final luminosity profile.

After fitting of the isophotes with ellipses, photometric and
morphological profiles have been obtained according to the procedure
described in Paper~I. For each galaxy, the
luminosity radial profile, the major axis position angle (defined 
from North to East), the ellipticity,
and $c4$ coefficient profiles as a function of the semi-major axis are
shown in Fig. 3.
The Fourier coefficient $c4$ measures the deviation of the isophotes
from the best fitting ellipse. A positive values indicate the isophote is
excessively elongated along the major axis, i.e., it is similar to a disk, 
while a negative $c4$ means the isophote is boxy. 

The residual background variations inside each frame were used to
derive, according to Fasano \& Bonoli (1990), proper errors for
morphological and photometric parameters.

For the dumbbell system in Fig. 3 we show only the luminosity profile
of the radio source.

\section{Comparison with previous results}

Several objects presented here were previously studied by Lilly
\& Prestage (1987,~LP87) and Smith \& Heckman (1989a,b,~SH89). In
common between our and LP87 samples there are 13 objects 
($0255+058$, $0325+023$, $0427-539$, 
$0430+052$, $0453-206$, $0620-526$, $0625-354$, $0625-536$, $0915-118$, 
$0945+076$, $1318-434$, $1333-337$, and $2221-023$), while 
eight are in common with SH89 ($0255+058$, $0325+023$, $0430+052$, 
$0945+076$, $1251-122$, $1717-009$, $1949+023$, and $2221-023$).

LP87 give Cousins metric R magnitudes for a fixed aperture of 19.2 kpc 
(for H$_0$=50~km sec$^{-1}$ Mpc$^{-1}$), whereas SH89 report V and B bands 
isophotal ($m_{25}$) magnitudes.
In order to perform an external check on our photometry, we derived
metric $R$ magnitudes at 19.2~kpc and isophotal magnitudes
$V_{25}$ (assuming $V-R=0.6$ as appropriate for low redshift elliptical galaxies) 
for the common objects (Fig. 4), finding on average:

$<\Delta R_{19.2}>_{LP87} = -0.12$ mag $~ (r.m.s = 0.29)$

$<\Delta V_{25}>_{SH89} = 0.02$ mag $~ (r.m.s = 0.36)$.

It is worth noticing that $0255+058$ and $1251-122$ are dumbbell
galaxies, while a bright, edge-on spiral galaxy projects on-top of $1318-434$. 
The measure of the luminosity of these galaxies is therefore particularly
difficult and dependent on the details of the adopted measuring strategy.
Not surprisingly $0255+058$ is the object with the largest discrepancy with
respect to SH89.
 
If we remove this object, the scatter becomes 0.27 and 0.19 for
the comparison with LP87 and SH89, respectively.
As a whole, our photometry agrees on average with previous photometry 
within $\sim 0.1$ magnitudes.

\section{Conclusions}

We presented surface photometry analysis for 50 radio galaxies which
complement our previous study of 29 objects from a complete sample of 95
low redshift 
radio galaxies.  Detailed  morphological and photometrical
properties of the galaxies are reported.  
As previously found by other studies (e.g., 
SH89; Ledlow \& Owen 1995), we confirm that the
galaxies 
associated with this kind of radio emission have mostly elliptical
morphology.

In some cases, however, the galaxy hosting the radio source is found
to have a substantial disc component (S0--like).
Moreover the surface brightness radial 
profiles of radio galaxies often exhibit deviations from 
de Vaucouleurs $r^{1/4}$ profile, 
due to the presence of nuclear point sources (likely 
associated with the active nucleus) and/or to 
low surface brightness extended halos.

A detailed description of each object is given in the appendix, while
a full discussion of the results for the whole observed sample of 79
radio galaxies is presented in Govoni et al. (1999).

\section*{Acknowledgments} 

This work was partly supported by the Italian Ministry for University and
Research (MURST) under grant Cofin98-02-32, and 
has made use of the NASA/IPAC Extragalactic Database (NED) which is
operated by the Jet Propulsion Laboratory, California Institute of 
Technology, under contract with the National Aeronautics and Space
Administration. 
 
%\Appendix{Results for individual objects}
%\appendix{\bf Appendix}
\appendix{}
\section{Results for individual objects}

${\bf 0255+058 :}$ The optical counterpart of 3C 75, the central radio
source of the cluster Abell 400, is an interesting case of a dumbbell
galaxy, with components separated by $\sim20\arcsec$ and with a radial
velocity difference of $500$ km/s. Twin radio jets depart from each of
the two optical nuclei (Owen et al. 1985), making this radio source
extremely unusual. The large scale radio structure is classified as 
FRI (Morganti et al. 1993). 

Unfortunately, a bright 
satellite crossed the field, passing close to the central
part of the dumbbell system (see Fig. 1).  Nevertheless, using the
AIAP masking facility, we were able to obtain a photometric
deblending, by iterative modeling of the two components.  In spite of
the appearance, the geometry of the two galaxies looks rather regular,
apart from a slight off-centering of the outermost
isophotes. Luminosity and geometrical profiles suggest 
both components are ellipticals. In particular, the luminosity profile
(see Fig. 3) of the northern galaxy is consistent with the presence
of a nuclear point source.

${\bf 0257-398 :}$ The optical counterpart of this radio source is
a rather isolated galaxy at $z\simeq 0.066$ (Scarpa et al. 1996).
The geometrical profiles of this object are highly suggestive of an S0
morphology (increasing ellipticity, positive $c4$ coefficient).

${\bf 0307-305 :}$ 
This object seems to be a rather isolated galaxy. Its luminosity and
geometrical profiles suggest a regular elliptical morphology, with a
possible nuclear point source.

${\bf 0312-343 :}$ The source looks like a large
elliptical with regular morphology. The luminosity profile
strongly suggests the presence of a bright nuclear point source.

${\bf 0325+023 :}$  The radio morphology of 3C 88 is dominated by
two symmetric and well developed lobes at PA$=60^\circ$, and is
classified as FRII (Morganti et al. 1993).
The optical counterpart which coincides with the radio nucleus, is an
elliptical galaxy, whose major axis has
position angle PA$=-30^\circ$. This is remarkably well defined
and stable (Fig. 3) and, within the uncertainty, orthogonal to
the radio structure.

${\bf 0332-391 :}$ A radio map of this FRI radio source is reported by
Jones \& McAdam (1992). Fig. 1 shows that the host galaxy is embedded 
in a quite dense environment. In particular, it seems to interact 
with another (very similar) galaxy, whose angular separation
from the radio galaxy is $\sim 76^{\prime\prime}$. The luminosity profile 
is consistent with the presence of a nuclear point source.

${\bf 0344-345 :}$ As for the previous case, the galaxy hosting this
radio source turns out to be embedded in a rich environment and seems
to interact with a close elliptical companion. The luminosity
profile is consistent with the presence of a nuclear point source,
whereas the geometrical profiles suggest regular elliptical
morphology. This source is also interesting  for having strong optical 
emission lines (Scarpa et al. 1996), and complex radio morphology 
(Jones \& McAdam 1992).

${\bf 0427-539 :}$  The optical counterpart of this radio source
is a spectacular case of dumbbell morphology in rich environment, with 
nuclei separated by $\sim 30^{\prime\prime}$ ($\sim$ 33 kpc).
The radio source is associated with the South-East component
of the dumbbell system. After deblending the two galaxies,
we observe a regular elliptical morphology for the brightest
object, whereas the other galaxy shows a rather amorphous and broad
light distribution (see contours in Fig. 2).

${\bf 0430+052 :}$ 3C 120 is a well studied radio source,
displaying superluminal motion (Zensus 1989). The spectrum of the optical counterpart
is quasar like (Tadhunter et al. 1993), and the galaxy has been often
classified as Seyfert 1, even if its spiral morphology has never been
clearly established.  

${\bf 0434-225 :}$ The galaxy hosting this radio source is embedded
in a moderately rich environment. It has regular elliptical
morphology, as illustrated by the luminosity and geometrical profiles,
as well as by its isophotal contours. The luminosity
profile is consistent with the presence of a nuclear point source.

${\bf 0446-206 :}$ The optical counterpart of this radio source
lies in the very dense environment of the cluster Abell~514.
The host galaxy is a normal elliptical.

${\bf 0452-190 :}$ An interesting case of dumbbell galaxy.
The radio source coincides with the Southernmost  component.
After deblending, both galaxies show very regular morphology
(see Fig. 2), suggesting the dumbbell appearance may just be
due to chance projection.  Based on the shape of their radial profile
we conclude the northernmost galaxy is an S0, and the other one
an elliptical.

${\bf 0453-206 :}$ In the optical band NGC 1692 looks like a large,
undisturbed elliptical galaxy, in spite of the presence of two nearby
galaxies both at a projected distance of $\sim
66^{\prime\prime}.5$ from the radio galaxy.  The south-east companion
is likely to be an edge-on spiral showing a pronounced C-shape.

${\bf 0511-305 :}$ Even if the optical counterpart of this radio source
 is located in a rather poor environment, the host galaxy
appears morphologically disturbed by the presence of some small
companions. In particular, two small compact galaxies on opposite
sides with respect to the galaxy center are aligned with an elongated 
structure extending for $\sim30^{\prime\prime}$ westward.

${\bf 0533-377 :}$ The optical counterpart of this radio source 
is an elliptical galaxy, located at the end of a chain of small galaxies. 
The surrounding environment is very dense, with several galaxies 
superposed on the main object.

${\bf 0546-329 :}$ The host galaxy of this radio source is
a large, normal elliptical. The only noticeable thing
being the presence of a nearby galaxy pair. 

${\bf 0548-317 :}$ A rather normal elliptical galaxy in a rich environment,
with several small galaxies  projecting on its halo. The luminosity profile 
suggests the  existence of a nuclear point source. 

${\bf 0620-526 :}$ This FRI radio source (Jones \& McAdam 1992) shows
weak emission lines superposed onto the continuum spectrum of a typical
early-type galaxy (Simpson et al. 1996). The source was also detected in
the X-ray band (Gioia \& Luppino 1994). Unfortunately, our image of
the galaxy is disturbed by the presence of several saturated columns of
the CCD, due to a nearby bright star. Nevertheless, using
appropriate masking we were able to produce a reliable radial profile, 
from which we infer the existence of a nuclear point source.

${\bf 0625-354 :}$ The optical counterpart of this FRI radio source,
located at the center of the cluster Abell 3392, is 
a giant elliptical embedded in a rich environment .
It is worth noticing the presence of a strong point source in the
nucleus of this galaxy, also supported by the emission lines 
observed in its optical spectrum (Tadhunter et al. 1993).

${\bf 0625-536 :}$ The optical counterpart of this radio
source is the South-East member of a dumbbell system,
located in the cluster Abell~3391.
After deblending the two galaxies with the iterative two-galaxy 
fitting procedure (Fig. 2), we found both galaxies show 
strong displacement of the isophotal centers roughly perpendicular to
their alignment, as expected in strong interactions.

${\bf 0634-205 :}$ The host galaxy is a normal elliptical lying in 
a rather poor environment. Strong emission lines
have been observed in its optical spectrum (Simpson et al. 1996).
The image of the galaxy is disturbed by the light from a satellite which
passed close to the center of the source. This structure was masked 
during isophotal analysis. 

${\bf 0712-349 :}$ The galaxy hosting this radio source is
a normal elliptical with regular morphology in a poor environment.

${\bf 0718-340 :}$ The optical counterpart looks like a normal 
elliptical with regular morphology in a poor environment.

${\bf 0806-103 :}$ 3C195 is an FRII radio source (Morganti et al. 1993),
with the optical counterpart exhibiting emission lines in its optical spectrum
(di Serego Alighieri et al. 1994). The host galaxy is located nearby a very
bright star which makes it difficult to derive luminosity and geometrical
profiles extended to the outer regions. 
The brighter part of the galaxy, which can be
reliably studied, suggests this galaxy has complex structure and
a nuclear point source. The
environment is relatively rich and at least two small galaxies may be
gravitationally interacting with the radio galaxy.

${\bf 0915-118 :}$ The optical counterpart of the FRI radio source
3C218 (Morganti et al. 1993), is an early-type galaxy 
most likely a member of a small group. Ionization emission 
lines were detected in its optical spectrum (Simpson et al. 1996).

${\bf 0940-304 :}$ The optical spectrum of this radio source is
of an early-type galaxy with old stellar population (Scarpa et al. 1996).
The optical morphology confirms this classification. The environment is poor.

${\bf 0945+076 :}$ The radio morphology of 3C227 is very elongated East-West
with terminal hot spots, and is classified as FRII (Morganti et al. 1993).
The optical counterpart resides in a poor environment, and
apart from the slight tendency to have disky
isophotes, the most important optical feature is the
presence of a very bright nuclear source. This is consistent with the
detection of strong emission lines with broad
wings in the optical spectrum (Simpson et al. 1996). The major axis of the 
galaxy, at almost constant position angle PA$\sim 0^\circ$, is perpendicular
to the radio structure. 

${\bf 1002-320 :}$ The optical counterpart is an elliptical galaxy,
most likely interacting with a nearby companion 
at projected distance $\sim 56^{\arcsec}$.

${\bf 1043-290 :}$ Its optical spectrum is
characteristic of an elliptical galaxy (Scarpa et al. 1996).
The surface photometry of this
galaxy is perturbed by the influence of a bright nearby star. Its
morphology is clearly of elliptical type and its luminosity profile is
reliable enough to indicate the presence of an outer halo. The
environment looks relatively rich.

${\bf 1053-282 :}$ This object lies in rich environment and is surrounded 
by several small galaxies, some of which are likely to be in 
interaction with the radio source. The surface
photometry is made difficult both because of the above mentioned
companions and also due to the presence of a relatively bright star
projecting onto the galaxy
body. Both morphology and luminosity profile
are consistent with the presence of a extended disk.

${\bf 1056-360 :}$ The radio structure of this source is large and
complex (Jones \& McAdam 1992). The optical
counterpart lies in a rich environment and coincides with a close galaxy
pair similar to $0452-190$.
The radio galaxy is the brightest of the two and has
normal luminosity and geometrical profiles for an elliptical galaxy. 
The optical spectrum exhibits emission lines (Simpson et al. 1996).

${\bf 1107-372 :}$ NGC 3557 is an FRI radio source (Birkinshaw \&
Davies 1985) in which H$\alpha$+[NII] extended emission has been
detected (Goudfrooij et al. 1994). The luminosity and geometrical
profiles indicate a regular elliptical morphology.

${\bf 1123-351 :}$ Its optical counterpart is a large elliptical
galaxy with undisturbed morphology.

${\bf 1251-122 :}$ 3C 278 is an FRI radio source (Morganti et al. 1993).
The optical counterpart is the Southernmost component of a dumbbell system.
After deblending the image of the two galaxies, we found
both components are heavily disturbed by tidal interaction.

${\bf 1251-289 :}$ The host galaxy inhabits a relatively poor
environment and shows undisturbed elliptical morphology in its outer
part. On the contrary, the inner region appears irregular and
elongated suggesting the presence of either a nuclear dust lane or a
double nucleus.

${\bf 1257-253 :}$  An elliptical galaxy with undisturbed morphology,
surrounded by several nearby companions. The optical 
spectrum is typical of early type galaxies without emission lines 
(Scarpa et al. 1996).

${\bf 1258-321 :}$ This radio galaxy lies in the cluster Abell 3537.
After masking the light from two nearby bright stars, the
surface photometry indicates a regular elliptical morphology and
a nuclear point source. 

${\bf 1318-434 :}$ NGC 5090 hosts an FRI radio source 
(Morganti et al. 1993) studied in detail by Lloyd et al. (1996). 
The galaxy looks like a normal giant elliptical, over which projects
NGC 5091, an edge-on spiral.
 
${\bf 1323-271 :}$ The host galaxy lies in cluster Abell 1736 and looks
like a normal elliptical in interaction with a small S0 galaxy.

${\bf 1333-337 :}$ A regular elliptical galaxy located at the 
center of cluster Abell 3565.
Optical emission lines were detected within few arcsec of the 
nucleus (Goudfrooij et al. 1994). 

${\bf 1344-241 :}$ The host galaxy lies in a rather poor environment.
Isophotal contours and geometrical profiles (increasing ellipticity,
constant position angle and positive $c4$ coefficient) 
suggest it is an S0 galaxy. However, signatures of a disc 
are not seen in the luminosity profile.

${\bf 1354-251 :}$ This radio galaxy inhabits a relatively poor
environment. Its optical morphology is of an undisturbed elliptical
galaxy, as confirmed also by the optical spectrum (Scarpa et al. 1996).

${\bf 1400-337 :}$ The optical counterpart is a large, regular
elliptical with a bright nucleus, located at the center of the cluster 
Abell~753.

${\bf 1404-267 :}$ A large and undisturbed elliptical galaxy lying at 
the center of the Abell cluster 3581. Since no radio maps are available  
for this radio source, we guess its FRI radio morphology from the
power-morphology relationship.

${\bf 1637-771 :}$ The radio source is estimated by Morganti et al. (1993)
to have an FRII 
radio morphology. Optical spectra have revealed the
presence of relatively strong emission lines (Tadhunter et al. 1993;
Simpson et al. 1996). The luminosity profile of the host galaxy is
consistent with the presence of a nuclear point source. The
environment is rather poor.

${\bf 1717-009 :}$ According to Morganti et al. (1993), 3C 353 is an FRII 
radio source. Emission lines have been detected in its optical spectrum 
(Tadhunter et al. 1993), and a LINER type spectrum has been also 
observed (Simpson et al. 1996). In our image the galaxy appears undisturbed
in the outer regions, while in the center both luminosity
and geometrical profiles support the existence of a bright point source.

${\bf 1733-565 :}$ An FRII  radio source (Morganti et al. 1993), with
emission lines in the optical spectrum 
(Simpson et al. 1996). The optical counterpart is most probably an
elliptical as suggested by its radial profile which precisely follows a 
de Vaucouleurs law. A  nuclear point source is also observed.

${\bf 1928-340 :}$ The host galaxy looks like a
normal elliptical located in a small galaxy group.

${\bf 1949+023 :}$ 3C 403 is an FRII radio source (Morganti et al. 1993) 
located in a poor environment. Strong emission lines were observed
in its optical spectrum (Simpson et al. 1996; Tadhunter et al. 1993).
In spite of the presence 
of a nearby bright star we derive for this galaxy a 
reliable profile, which clearly indicates a normal elliptical structure. 

${\bf 2221-023 :}$ 3C~445 is a well known radio source, and broad
emission lines have been observed in its optical spectrum (Eracleous \& 
Halpern 1994; Corbett et al. 1998). The optical morphology is characterized 
by the presence of an extremely bright nuclear point source.\\
{\bf References.}

\ref {Birkinshaw M., Davies R.L., 1985, ApJ 291, 32}

\ref {Colina L., de Juan L., 1995, ApJ 448, 548}

\ref {Corbett E.A., Robinson A., Axon D.J., Young S., Hough J.H., 1998,
MNRAS 296, 721}

\ref {de Juan L., Colina L., Perez-Fournon I., 1994, ApJS 91, 507}

\ref {di Serego Alighieri S., Danziger I.J., Morganti R., Tadhunter
C.N., 1994, MNRAS 269, 998}

\ref {Ekers R.D., Wall J.V., Shaver P.A., et al., 1989, MNRAS 236, 737 (EK)}

\ref {Eracleous M., Halpern J.P., 1994, ApJS 90, 1}

\ref {Fanaroff B.L., Riley J.M., 1974, MNRAS 167, 31}

\ref {Fasano G., 1990, internal report of Astr. Obs. of Padova}

\ref {Fasano G., Bonoli C., 1990, A\&A 234, 89}

\ref {Fasano G., Falomo R., Scarpa R., 1996, MNRAS 282, 40 (Paper I)}

\ref {Gioia I.M., Luppino G.A., 1994, ApJS 94, 583} 

\ref  {Gonzalez-Serrano J.I., Carballo R., Perez-Fournon I., 1993, AJ 105, 1710}

\ref {Goudfrooij P., Hansen L., Jorgensen H.E., Norgaard-Nielsen H.U., 1994, A\&AS 105, 341}

\ref {Govoni F., Falomo R., Fasano G., Scarpa R., 1999, A\&A, submitted}

\ref {Hine R.G, Longair M.S., 1979, MNRAS 188, 111}

\ref {Jones P.A., McAdam W.B., 1992, ApJS 80, 137}

\ref {Knapp G.R., Kerr F.J., 1974, A\&A 35, 361}

\ref {Landolt A.U., 1992, AJ 104, 340}

\ref {Ledlow M.J., Owen F.N., 1995, AJ 109, 853 }

\ref {Lilly S.J., Prestage R.M., 1987, MNRAS 225, 531 (LP87)}

\ref {Lloyd B.D., Jones P.A., Haynes R.F., 1996, MNRAS 279, 1197}

\ref {Longair M.S., Seldner M., 1979, MNRAS 189, 433}

\ref {Morganti R., Killeen N.E.B., Tadhunter C.N., 1993, MNRAS 263, 1023}

\ref {Morganti R., Oosterloo T.A., Reynolds J.E., Tadhunter C.N.,
 Migenes V., 1997, MNRAS 284, 541}

\ref {Owen F.N., Laing R.A., 1989, MNRAS 238, 357}

\ref {Owen F.N., White R.A., 1991, MNRAS 249, 164}

\ref {Owen F.N., Odea C.P., Inoue M., Eilek J.A., 1985, ApJ 294, L85}

\ref {Prestage R.M., Peacock J.A., 1988, MNRAS 230, 131}

\ref {Scarpa R., Falomo R., Pesce J.E., 1996, A\&AS 116, 295}

\ref {Simpson C., Ward M., Clements D.L., Rawlings S., 1996, MNRAS
      281, 509}

\ref {Smith E.P., Heckman T.M., 1989a, ApJS 69, 365 (SH89)}

\ref {Smith E.P., Heckman T.M., 1989b, ApJ 341, 658 (SH89)}

\ref {Stark A.A., Gammie C.F., Wilson R.W., et al., 1992, ApJS 79, 77}

\ref {Tadhunter C.N., Morganti R., Di Serego Alighieri S.,
Fosbury R.A.E., Danziger I.J., 1993, MNRAS 263, 999}

\ref {Wall J.V., Peacock J.A., 1985, MNRAS 216, 173 (WP)}

\ref {Zensus J.A., 1989, in BL Lac objects, ed. L. Maraschi,
T. Maccacaro, M.H. Ulrich (Berlin: Springer), p.3}

\end{document}